\documentclass{article}
\usepackage{spconfa4,amsmath,graphicx}
\usepackage{hyperref}
\usepackage{booktabs}
\usepackage{tabularx}
\usepackage{multirow}
\usepackage{amsmath}
\usepackage{amssymb}
\usepackage{mathtools}
\usepackage[number-mode=math]{siunitx}

\newcommand{\VEC}[1]{\mathbf{#1}}          

\newcommand{\putindex}[3]{\vtop{\hbox{\hspace{#3} $#1$}
            \hbox{\raise 6mm \hbox{$\scriptscriptstyle #2$}}}}

\newcommand{\gradx}[0]{\vtop{\hbox{\rm grad}
            \hbox{\raise 2.5mm \hbox{\rm \hspace{2mm} \footnotesize x}}}}

\newcommand{\grady}[0]{\vtop{\hbox{\rm grad}
            \hbox{\raise 2.5mm \hbox{\rm \hspace{2mm} \footnotesize y}}}}

\newcommand{\grad}[1]{\vtop{\hbox{\rm grad}
            \hbox{\raise 2.5mm \hbox{#1}}}}

\newcommand{\btb}{     \begin{tabbing}             }
\newcommand{\bte}{     \end{tabbing}               }

\usepackage[mode=buildnew]{standalone}
\usepackage{xcolor}
\usepackage{tikz}
\usetikzlibrary{chains}
\usetikzlibrary{shapes.geometric, arrows.meta, positioning, calc}

\definecolor{condred}{RGB}{176,0,70}
\definecolor{scoreblue}{RGB}{0,128,180}
\providecolor{cBlack}{RGB}{0,0,0}
\providecolor{cGreen}{RGB}{32,128,64}
\providecolor{cDarkRed}{RGB}{128,48,32}
\providecolor{cBlue}{RGB}{32,64,128}

\title{DiffVQE2: An Efficient Low-delay Diffusion Model \linebreak for Acoustic Echo and Noise Control}
\name{Haljan Lugo$^*$, Ernst Seidel$^*$, Pejman Mowlaee$^\circ$, Ziyue Zhao$^\circ$, Tim Fingscheidt$^*$}
\address{
    $^*$Institute for Communications Technology, Technische Universität Braunschweig, \\
    Schleinitzstraße~22, 38106~Braunschweig, Germany \\
    $^\circ$ GN Advanced Science, Lautrupbjerg 7, 2750~Ballerup, Denmark}

\begin{document}
\ninept
\maketitle
\begin{abstract}
Hands-free communication devices and speakerphones are inherently affected by acoustic echo and background noise.
To mitigate these impairments, end-to-end discriminatively trained neural networks have emerged as the best-performing approach in research and deployment.
While recent advancements in generative methods have provided remarkable results for various speech enhancement tasks, diffusion-based acoustic echo control (AEC) research is still restricted to non-causal, utterance-level processing, thereby not widely applicable in practice.
With this work, we are the first to propose low-delay (i.e., causal) diffusion-based joint AEC and noise control models \texttt{DiffVQE2} / \texttt{DiffVQE2-S}, excelling the so-far state of the art \texttt{DeepVQE} / \texttt{DeepVQE-S} models in multiple objective metrics, and, most importantly, in subjective MOS, respectively. In addition, our models are less complex. Furthermore, we show that a limited lookahead applied to the efficient \texttt{DiffVQE2-S} model allows for an even higher performance. These results have been obtained on the ICASSP 2023 AEC Challenge blind test set.
We claim the first streaming-capable, diffusion-based acoustic echo and noise control 
that excels state-of-the-art discriminative approaches.

\end{abstract}
\section{Introduction}
Modern speech enhancement for edge devices such as mobile phones and speakerphones has traditionally relied on compact, discriminatively trained deep neural networks (DNNs), optimized to directly map corrupted speech to a clean estimate.
For hands-free communication, acoustic echo control (AEC) is among the most critical of these tasks, suppressing the acoustic feedback path between loudspeaker and microphone that degrades full-duplex voice quality.
With the widespread adoption of generative modeling across machine learning, recent speech enhancement research has begun shifting towards probabilistic and generative techniques.
Diffusion-based models in particular have demonstrated remarkable performance on noise reduction by learning the clean speech data distribution and sampling from it during inference~\cite{Welker2022,Richter2023,Lemercier2022,Scheibler2024,Fu2025}, consistently surpassing discriminative baselines on perceptual quality metrics.
Results from the Interspeech 2025 URGENT Challenge~\cite{Saijo2025} further underscore the potential of these techniques across a broad range of distortion types, motivating their extension to the more constrained AEC setting.

In recent years, AEC has been primarily tackled by discriminative models such as Microsoft's \texttt{DeepVQE} and its compact variant \texttt{DeepVQE-S}~\cite{Indenbom2023}, which has shown strong performance and provides a streaming-capable solution already being deployed.
On ICASSP 2023 AEC Challenge~\cite{Cutler2024} data, \texttt{DeepVQE} outperformed all challenge entries, establishing a clear reference method for what constitutes a well-performing joint acoustic echo and noise control model.
Only recently, generative modeling has made a step into the AEC domain as well:
In \cite{Liu2024a}, a version of the \texttt{StoRM} framework~\cite{Lemercier2022} was adapted to the AEC task, however, lacking reproducible training procedures, data, and model definition.
This is addressed in \cite{Lugo2026}, where \texttt{DiffVQE} provides the first hybrid diffusion-based AEC and noise reduction model which is fully reproducible, building on the frameworks of \cite{Fu2025} and \cite{Seidel2024}, while using publicly available data from the Interspeech 2025 URGENT Challenge~\cite{Saijo2025}.
However, both generative approaches operate on utterance level, making them incompatible with the low-delay (i.e., causality) requirements of real-world hands-free communication systems.

In this work, we propose an efficient and low-delay, hybrid diffusion-based model for joint echo and noise control \texttt{DiffVQE2}, derived from the non-causal larger \texttt{DiffVQE}~\cite{Lugo2026}.
We do this by adapting the model to work in a causal fashion using causal convolutions and unidirectional instead of bidirectional gated recurrent units (GRUs), and by further introducing a smaller \texttt{DiffVQE2-S}, slimming the network to a similar model size and a lower computational complexity as \texttt{DeepVQE-S}.
To close the performance gap introduced by causal processing, we further introduce
a limited lookahead, allowing a trade-off between algorithmic delay and output quality by partially compensating for the absence of future context.
\textit{We thereby achieve state-of-the-art performance in the model regimes of both \texttt{DeepVQE} and \texttt{DeepVQE-S} on the ICASSP 2023 AEC Challenge test set.
To the best of our knowledge, we provide the first generative, streaming-capable model for joint acoustic echo and noise control.}

\section{Methods}

\subsection{Score-based Diffusion for Voice Quality Enhancement}
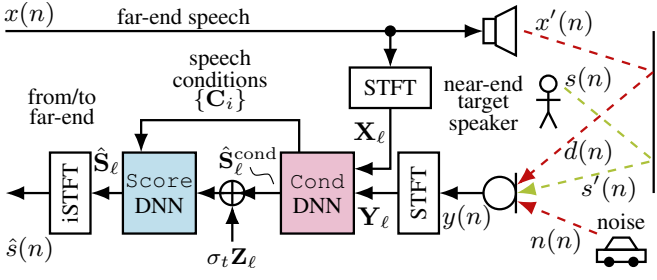
\begin{figure}[!t]
    \centering
\definecolor{lightred}{RGB}{235,191,209}%
\definecolor{lightblue}{RGB}{191,223,236}%
\definecolor{audiogreen}{RGB}{172,193,58}%
\definecolor{audiored}{RGB}{190,15,15}%
\begin{tikzpicture}[
    >=Latex,
    thick,     
    font=\footnotesize,
    inner sep=0pt,
    outer sep=0pt,
    block/.style={
        draw, rectangle, minimum height=0.5cm, minimum width=1.0cm, 
        align=center, fill=white
    },
    concat/.style={
        draw, rectangle, minimum height=2.0em, minimum width=2.5em,
        align=center, fill=lightorange
    },
    acoustic/.style={
        dashed, draw=gray, thick, ->
    },
    pics/person/.style={
        code={
            \draw[thick] (0,0.35) circle (0.1); 
            \draw[thick] (0,0.25) -- (0,0);  
            \draw[thick] (-0.15,0.15) -- (0.15,0.15); 
            \draw[thick] (0,0) -- (-0.1,-0.2); 
            \draw[thick] (0,0) -- (0.1,-0.2);  
        }
    },
    pics/speaker/.style={
        code={
            \draw[thick, fill=white] (0,0) rectangle (0.15,0.3);
            \draw[thick, fill=white] (0.15,-0.025) -- (0.4,-0.15) -- (0.4,0.45) -- (0.15,0.325) -- cycle;
        }
    },
    pics/mic/.style={
        code={
            \draw[thick, fill=white] (0, 0) circle (0.2);
            \draw[thick] (0.2,-0.25) -- (0.2, 0.25);
        }
    },
    pics/car/.style={
        code={
            \draw[thick] (0,0) -- (0.1,0.15) -- (0.3,0.15) -- (0.4,0) -- cycle;
            \draw[thick] (-0.1,0) rectangle (0.5, -0.15);
            \draw[fill=black] (0.05,-0.15) circle (0.05);
            \draw[fill=black] (0.35,-0.15) circle (0.05);
        }
    },
    adder_geo/.style={
        draw, circle, 
        minimum size=0.3cm, 
        fill=white,
        path picture={
            \draw[black] (path picture bounding box.south) -- (path picture bounding box.north);
            \draw[black] (path picture bounding box.west) -- (path picture bounding box.east);
    }
},
]


\draw[thick] (8.0, 1.0) -- (8.0, 3.0); 

\path (5.9, 2.85) pic (spk) {speaker};
\path (6.1, 1) pic (mic) {mic};
\coordinate (ne_spk) at (6.7, 2.0);
\path (ne_spk) pic (interferer) {person};
\coordinate (noise_pos) at (7.3, 0.5);
\path ($(noise_pos) + (0.1, -0.2)$) pic (noise) {car};

\node at (2.2,3.15) {far-end speech};

\node at ($(ne_spk) + (-0.8,0.35)$) {near-end};
\node at ($(ne_spk) + (-0.8,0.1)$) {target};
\node at ($(ne_spk) + (-0.8,-0.15)$) {speaker};
\node at ($(ne_spk) + (0.5,0.4)$) {\begin{small}{$s(n)$}\end{small}};
\node at (7.45, 1.05) {\begin{small}{$s'(n)$}\end{small}};
\node at (0.3, 3.2) {\begin{small}{$x(n)$}\end{small}};
\node at (6.9, 3.1) {\begin{small}{$x'(n)$}\end{small}};
\node at (5.7, 0.7) {\begin{small}{$y(n)$}\end{small}};
\node at (7.2, 1.5) {\begin{small}{$d(n)$}\end{small}};
\node at ($(noise_pos) + (-0.5, -0.1)$) {\begin{small}{$n(n)$}\end{small}};
\node at ($(noise_pos) + (0.3, 0.15)$) {noise};


\coordinate (source_pos) at (0.7, 2.1); 
\node[above=0.0cm of source_pos] {from/to};
\node[above=-0.25cm of source_pos] {far-end};

\draw[->] (0, 3.0) -- (5.9, 3.0);

\fill (4.75, 3.0) circle (2pt) coordinate (tap_point);

\node[block] (stft_top) at (4.75, 2.3) {STFT};
\node at (4.5, 1.75) {\begin{small}{$\VEC{X}_\ell$}\end{small}};

\node[block, rotate=90] (stft_bot) at (5.1, 1) {STFT};
\node at (4.55, 0.7) {\begin{small}{$\VEC{Y}_\ell$}\end{small}};


\coordinate (cond_center) at (3.85, 1);

\def\cw{0.45}  
\def\ch{0.5}  
\def\th{0.5}  
\def\tw{0.0}  

\draw[thick, fill=lightred] 
    ($(cond_center) + (-\cw, \ch)$) --   
    ($(cond_center) + (-\tw, \th)$) --   
    ($(cond_center) + (\tw, \th)$) --    
    ($(cond_center) + (\cw, \ch)$) --    
    ($(cond_center) + (\cw, -\ch)$) --   
    ($(cond_center) + (\tw, -\th)$) --   
    ($(cond_center) + (-\tw, -\th)$) --  
    ($(cond_center) + (-\cw, -\ch)$) --  
    cycle;

\node at ($(cond_center) + (0, 0.15)$) {\texttt{Cond}}; 
\node at ($(cond_center) + (0, -0.15)$) {DNN}; 

\coordinate (cond_east) at ($(cond_center) + (\cw, 0)$);
\coordinate (cond_west) at ($(cond_center) + (-\cw, 0)$);

\node[adder_geo] (my_sum) at (2.8, 1) {}; 

\coordinate (score_center) at (1.9, 1);

\def\cws{0.45};  
\def\chs{0.5};  
\def\ths{0.5};  
\def\tws{0.0};  

\draw[thick, fill=lightblue] 
    ($(score_center) + (-\cws, \chs)$) --   
    ($(score_center) + (-\tws, \ths)$) --   
    ($(score_center) + (\tws, \ths)$) --    
    ($(score_center) + (\cws, \chs)$) --    
    ($(score_center) + (\cws, -\chs)$) --   
    ($(score_center) + (\tws, -\ths)$) --   
    ($(score_center) + (-\tws, -\ths)$) --  
    ($(score_center) + (-\cws, -\chs)$) --  
    cycle;

\node at ($(score_center) + (0, 0.15)$) {\texttt{Score}};
\node at ($(score_center) + (0, -0.15)$) {DNN};

\coordinate (score_east) at ($(score_center) + (\cws, 0)$);
\coordinate (score_west) at ($(score_center) + (-\cws, 0)$);

\node[block, rotate=90] (istft) at (0.8, 1) {iSTFT};
\node at (0.3, 0.3) {\begin{small}{$\hat{s}(n)$}\end{small}};
\node at (1.25, 1.4) {\begin{small}{$\hat{\VEC{S}}_\ell$}\end{small}};


\draw[->] (tap_point) -- (stft_top.north);


\draw[->] (stft_top.south) -- (4.75, 1.3) -- ($(cond_east) + (0, 0.3)$);

\draw[->] (5.9, 1) -- (stft_bot.south);


\draw[->] (stft_bot.north) -- (cond_east);


\node at (3.0, 1.4) {$\hat{\VEC{S}}^{\mathrm{cond}}_\ell$};
\draw[line width=0.4pt] plot[smooth] coordinates { (3.02,1.32) (3.05,1.25) (3.10,1.22) (3.25,1.23) (3.31,1.11) };

\draw[->] (cond_west) -- ++(-0.55, 0); 
\draw[->] (2.7, 1) -- (score_east); 

\node at (2.8, 0.2) {$\sigma_t\VEC{Z}_\ell$};
\draw[->] (2.8, 0.4) -- (2.8, 0.85);

\draw[->] ($(cond_center) + (-\cws/2-\tws/2, +\chs/2+\ths/2)$) -- ++(0, 0.4) -- ++(-1.95, 0) -- ($(score_center) + (-\cws/2-\tws/2, +\chs/2+\ths/2)$);
\node at (2.65, 2.65) {speech};
\node at (2.65, 2.4) {conditions};
\node at (2.65, 2.15) {$\{\VEC{C}_i\}$};

\draw[->] (score_west) -- (istft.south);

\draw[->] (istft.north) -- (0, 1);


\draw[acoustic, audiored] (6.4, 3.0) -- (8, 2.5) -- (6.31, 1.1);

\draw[acoustic, audiogreen] ($(ne_spk) + (0.2, 0.3)$) -- (8, 1.4) -- (6.31, 1.0);

\draw[acoustic, audiored] (noise_pos) -- (6.31, 0.9);
\end{tikzpicture}
    \caption{\textbf{Inference setup of \texttt{DiffVQE2}} using small, causal adaptions of \textcolor{condred}{\texttt{Cond}} and \textcolor{scoreblue}{\texttt{Score}} networks. Figure adapted from \cite{Lugo2026}.}
    \label{fig:blockdiagram}
\end{figure}

In \autoref{fig:blockdiagram}, we provide an overview of the inference setup of our \texttt{DiffVQE2} method in a hands-free system.
Far-end signal $x(n)$ is played back by the loudspeaker and subsequently distorted by loudspeaker nonlinearities ($x'(n)=f_{\mathrm{NL}}(x(n))$) and room reverberations resulting in an echo signal $d(n)=h_1(n)*x'(n)$.
Moreover, the near-end target speech $s(n)$ is also subject to reverberation $s'(n)=h_2(n)*s(n)$, with $h_1(n)$ and $h_2(n)$ denoting two room impulse responses which correspond to different (loud)speaker positions in the same room.
Together with background noise $n(n)$, we obtain the corresponding microphone signal by summation of all components $y(n)=s'(n)+d(n)+n(n)$.
The goal of echo and noise control is to leverage the knowledge about the loudspeaker signal $x(n)$ to enhance the microphone signal $y(n)$ such that one gets an estimate of the target clean speech $\hat{s}(n)$.
In our case, we do this by first transforming all signals into the frequency domain, where we process the signals frame-by-frame.
We denote $\VEC{X}_1^L=(\VEC{X_{\ell}})\in\mathbb{C}^{K\times L}$ as the reference signal, $\VEC{Y}_1^L=(\VEC{Y_{\ell}})\in\mathbb{C}^{K\times L}$ as the microphone signal, and $\VEC{S}_1^L=(\VEC{S_{\ell}})\in\mathbb{C}^{K\times L}$ as the clean speech signal spectrogram of frames $1$ to $L$, where $k\in\mathcal{K}=\{0,\dots,K\!-\!1\}$ is the frequency bin index.
Then we estimate a discriminative mask (resulting in an intermediate estimation $\hat{\VEC{S}}^{\mathrm{cond}}_\ell$) and a set of speech conditions $\{\VEC{C}_i\}$ using the conditioning DNN \texttt{Cond}.
Then, we add Gaussian noise $\sigma_t\VEC{Z}_\ell$ and do a one-step score matching-based generative estimation using the initial estimation $\hat{\VEC{S}}^{\mathrm{cond}}_\ell$, the speech conditions $\{\VEC{C}_i\}$, and the diffusion noise scale $\sigma_t$.
Finally, we transform the estimated signal back into the time domain to achieve the estimate $\hat{s}(n)$.

We adopt the score-based diffusion setup of~\cite{Lugo2026} without modification, which in turn builds on the stochastic differential equation (SDE) formulation of~\cite{Scheibler2024,Fu2025} with reverse-time dynamics from~\cite{Anderson1982}.
Training combines a denoising score matching objective $J^{\mathrm{SM}}$~\cite{Vincent2011} with the complex compressed reconstruction loss $J^{\mathrm{CC}}$ of ~\cite{Braun2020b} applied to the conditional estimates $\hat{\VEC{S}}^{\mathrm{cond}}_\ell$ and the final system outputs $\hat{\VEC{S}}_\ell$.
At inference time, we use the EDM-style~\cite{Karras2022} one-step sampler as in \cite{Fu2025,Lugo2026}, trading the iterative reverse process for a single score evaluation conditioned on $\{\VEC{C}_i\}$ and $\hat{\VEC{S}}^{\mathrm{cond}}_\ell$.

\subsection{Novel \texttt{DiffVQE2}}
The non-causal \texttt{DiffVQE} backbone (for details, see~\cite{Lugo2026}) cannot operate in a streaming fashion due to two immanent dependencies on future context: symmetric temporal convolutions that read ahead of the current frame, and bidirectional gated recurrent units (BGRUs) whose hidden states aggregate the full sequence before producing an output.
Both, the \texttt{Cond} DNN and \texttt{Score} DNN, share this general structure.
Thus, we apply our causal reformulation to both in an analogous fashion.
We replace all symmetric temporal convolutions with left-padded causal convolutions~\cite{Oord2016}.
This ensures that the receptive field covers only past and current frames.
Moreover, we replace BGRUs by standard unidirectional GRUs, preserving some recurrent capacity and increase the hidden size to recover the parameter and compute budget while eliminating the reverse pass.
We call the resulting low-delay (i.e., causal) network \texttt{DiffVQE2}.

As edge devices require small model footprints, both with regards to number of parameters, as well as computational complexity, we apply two changes to achieve this.
Starting from the causal \texttt{DiffVQE2}, we first reduce the encoder (and mirrored decoder) channel widths across all levels, thereby significantly reducing computational complexity.
Second, we replace the GRU layers with grouped GRUs~\cite{Tan2020} that partition the latent features into $g$ independent groups, processing each of these in parallel.
For a GRU with input size $n_\mathrm{in}$ and hidden size $n_\mathrm{hid}$, the parameter count is $n_\mathrm{GRU}=3\cdot(n_\mathrm{hid}^2 + n_\mathrm{in}\cdot n_\mathrm{hid} + 2n_\mathrm{hid})$.
Thus, grouping of input and hidden states reduces the parameter count by a factor of $\sim g\cdot1/g^2$, as $n_\mathrm{GRU}$ is dominated by the quadratic terms.
The resulting efficient low-delay network we call \texttt{DiffVQE2-S}.

To recover quality lost to strict causality requirements without sacrificing the streaming capability of \texttt{DiffVQE2-S}, we introduce a limited lookahead of $N_\mathrm{L}$ frames.
The \texttt{Cond} DNN buffers $N_\mathrm{L}$ future STFT frames before processing the current frame $(\VEC{Y}_\ell, \VEC{X}_\ell)$.
We therefore introduce further algorithmic delay of $N_\mathrm{L}$ times the frameshift.
To best incorporate this future information into robust feature representations, we unfold the spectrogram and concatenate the shifted version of the spectrogram in the channel dimension before feeding it into the first convolution of the \texttt{Cond} DNN.

\subsection{Training}
\begin{figure}[!t]
    \centering
    \definecolor{lightred}{RGB}{235,191,209}
\definecolor{lightblue}{RGB}{191,223,236}
\begin{tikzpicture}[
    >=Latex,
    thick,
    font=\footnotesize,
    inner sep=0pt,
    outer sep=0pt,
    condblock/.style={
        draw, rectangle, minimum height=0.8cm, minimum width=1.4cm,
        align=center, fill=lightred, line width=1pt
    },
    scoreblock/.style={
        draw, rectangle, minimum height=0.8cm, minimum width=1.4cm,
        align=center, fill=lightblue, line width=1pt
    },
    lossblock/.style={
        draw, rectangle, minimum height=0.5cm, minimum width=1.0cm,
        align=center, fill=white, line width=1pt
    },
    adder/.style={
        draw, circle, minimum size=0.3cm, fill=white, line width=1pt,
        path picture={
            \draw[black] (path picture bounding box.south) -- (path picture bounding box.north);
            \draw[black] (path picture bounding box.west) -- (path picture bounding box.east);
        }
    },
    citwopath/.style={dashed, line width=1pt},
]
\draw[white, thin] (0,0) rectangle (8.6,5.0);

\node[condblock]  (cond)    at (2.2,  4.0)  {\texttt{Cond}\\DNN};
\node[adder]      (addL)    at (2.2, 2.7)  {};
\node[adder]      (addR)    at (7.6, 2.7)  {};
\node[scoreblock] (scoreL)  at (2.2, 1.7)  {\texttt{Score}\\DNN};
\node[scoreblock] (scoreR)  at (7.6, 1.7)  {\texttt{Score}\\DNN};
\def\lossYheight{0.25};
\node[lossblock]  (lossCC1) at (0.9, \lossYheight) {$J^{\mathrm{CC}}$};
\node[lossblock]  (lossCC2) at (2.2, \lossYheight) {$J^{\mathrm{CC}}$};
\node[lossblock]  (lossSM)  at (7.6, \lossYheight) {$J^{\mathrm{SM}}$};

\def\inYheight{4.8};
\node[font=\footnotesize, anchor=west]  (Xinput) at (0.0, \inYheight) {$\VEC{X}_{1:\ell+N_{\mathrm{L}}}$};
\node[font=\footnotesize, anchor=west]  (Yinput) at (2.7, \inYheight) {$\VEC{Y}_{1:\ell+N_{\mathrm{L}}}$};

\draw[->]  (Xinput.east) -- (1.85, \inYheight) -- ($(cond.north)+(-0.35,0)$);

\draw[<-] (cond.north) -- (2.2, \inYheight) -- (Yinput.west);

\node[font=\footnotesize] (Sinput) at ($(addR.north)+(0, 1.35)$) {$\VEC{S}_{1:\ell}$};
\draw[->] ($(Sinput.south) + (0, -0.1)$) -- (addR.north);

\coordinate (condTap) at ($(cond.south)+(0,-0.3)$);
\draw[->] (cond.south) -- (addL.north);
\fill (condTap) circle (2pt);

\draw[->] (addL.south) -- (scoreL.north);
\draw[->] (addR.south) -- (scoreR.north);

\coordinate (sigSrc)      at (5.65, 2.7);
\node[font=\footnotesize] at ($(sigSrc)+(0, 1.5)$) {$\sigma_t\VEC{Z}_{1:\ell}$};


\node[font=\tiny] at ($(sigSrc)+(0.8, 0.1)$) {$B\!\times\!2\!\times\!K\!\times\!L$};
\node[font=\tiny] at ($(sigSrc)+(-2.3, 0.1)$) {$B\!\times\!2\!\times\!K\!\times\!L$};

\def\ciOffset{0.075};
\def\ciVx{4.9};

\coordinate (gaussTap) at (sigSrc);
\fill (gaussTap) circle (2pt);
\draw[->] (gaussTap.west) -- (addL.east);
\draw[->] (gaussTap.east) -- (addR.west);
\node[font=\footnotesize, anchor=north, align=center] at ($(gaussTap)+(0,-0.05)$) {Gaussian};
\node[font=\footnotesize, anchor=north, align=center] at ($(gaussTap)+(0,-0.3)$) {noise};

\draw[line width=1pt] ($(sigSrc)+(0, 1.15)$) circle (0.15cm);
\node[font=\footnotesize] at ($(sigSrc)+(0, 1.125)$) {$\sim$};
\draw[-] (gaussTap.north) -- ++(0, 1.0);

\coordinate (ciStartUp) at ($(cond.east)+(0.105,+\ciOffset)$);
\coordinate (ciStartDn) at ($(cond.east)+(0.105,-\ciOffset)$);

\coordinate (ciJunction) at (\ciVx, 1.9);
\fill (ciJunction) circle (2pt);

\draw[citwopath] (ciStartUp) -- ($(cond.east)+(0.7,+\ciOffset)$);
\draw[citwopath] (ciStartDn) -- ($(cond.east)+(0.7,-\ciOffset)$);

\draw[citwopath] (cond.east) -- ($(cond.east)+(2.0,0)$) -- (ciJunction);

\draw[citwopath] ($(scoreL.east)+(0.3125,0.2+\ciOffset)$) -- ($(ciJunction)+(-1.1,+\ciOffset)$);
\draw[citwopath] ($(scoreL.east)+(0.3125,0.2-\ciOffset)$) -- ($(ciJunction)+(-1.1,-\ciOffset)$);

\draw[-{Latex[width=0.3cm]}, line width=1pt, dashed] (ciJunction) -- ($(scoreL.east)+(0,0.2)$);

\draw[citwopath] ($(scoreR.west)+(-0.3125,0.2+\ciOffset)$) -- ($(ciJunction)+(1.1,+\ciOffset)$);
\draw[citwopath] ($(scoreR.west)+(-0.3125,0.2-\ciOffset)$) -- ($(ciJunction)+(1.1,-\ciOffset)$);

\draw[-{Latex[width=0.3cm]}, line width=1pt, dashed] (ciJunction) -- ($(scoreR.west)+(0,0.2)$);

\node[font=\tiny, anchor=west] at ($(cond.east)+(0.1, 0.25)$) {$B\!\times\!C_i\!\times\!K_i\!\times\!L$};
\node[font=\footnotesize, anchor=east] at (\ciVx-0.05, 3.375) {$\{\VEC{C}_i\}$};
\node[font=\footnotesize, anchor=east, align=right] at (\ciVx-0.75, 3.5) {speech};
\node[font=\footnotesize, anchor=east, align=right] at (\ciVx-0.75, 3.25) {conditions};


\draw[->, dotted, line width=1pt] (condTap) -- ++(-1.3, 0.0) -- (lossCC1.north);

\node[font=\tiny, anchor=east] at ($(condTap)+(-0.1, 0.1)$) {$B\!\times\!2\!\times\!K\!\times\!L$};

\node[anchor=east] at (0.8, \lossYheight+0.75) {$\hat{\VEC{S}}^{\mathrm{cond}}_{1:\ell}$};

\draw[->, dotted, line width=1pt] (scoreL.south) -- (lossCC2.north);
\node[anchor=east] at (2.1, \lossYheight+0.75) {$\hat{\VEC{S}}_{1:\ell}$};
\node[font=\tiny, anchor=west] at (2.3, \lossYheight+0.9) {$B\!\times\!2\!\times\!K\!\times\!L$};

\draw[->, dotted, line width=1pt] (scoreR.south) -- (lossSM.north);
\node[anchor=west] at (7.7, \lossYheight+0.75) {$\hat{\VEC{S}}_{t,1:\ell}$};
\node[font=\tiny, anchor=east] at (7.5, \lossYheight+0.9) {$B\!\times\!2\!\times\!K\!\times\!L$};

\draw[{Triangle[length=0.5cm, width=0.5cm]}-{Triangle[length=0.5cm, width=0.5cm]}, lightgray, line width=0.3cm] ($(scoreL.east)+(0,-0.2)$) -- ($(scoreR.west)+(0,-0.2)$);
\node[font=\footnotesize] at ($(scoreL.east)!0.5!(scoreR.west)+(0,-0.2)$) {shared weights};
\end{tikzpicture}
    \caption{\textbf{\texttt{DiffVQE2} training setup} with limited lookahead $N_\mathrm{L}$.
    The \textcolor{condred}{\texttt{Cond}} DNN receives $N_{\mathrm{L}}$ future frames as input but produces estimates up to frame $\ell$. The \textcolor{scoreblue}{\texttt{Score}} DNN is applied twice: to the noised conditional estimate (left) and the noised clean speech (right).}
    \label{fig:trainsetup}
\end{figure}
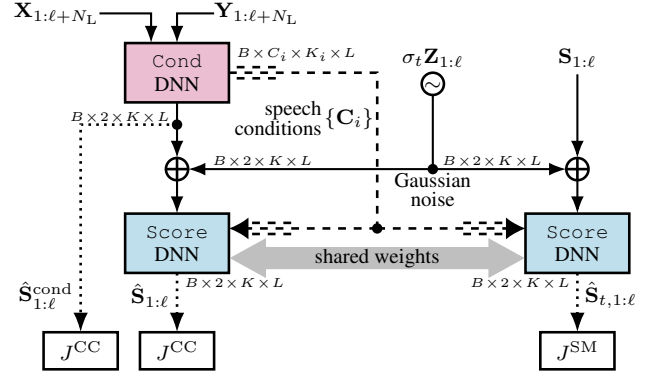

We train using the joint objective of~\cite{Lugo2026}, as depicted in \autoref{fig:trainsetup}:
\begin{equation}
    J = J^{\mathrm{CC}}(\hat{\VEC{S}}^{\mathrm{cond}}_{1:\ell}, \VEC{S}_{1:\ell})
    + J^{\mathrm{CC}}(\hat{\VEC{S}}_{1:\ell}, \VEC{S}_{1:\ell})
    + \alpha J^{\mathrm{SM}}\left(\hat{\VEC{S}}_{t,1:\ell}, \frac{\VEC{Z}_{1:\ell}}{\sigma_t}\right).
\end{equation}
The first term supervises the \texttt{Cond} DNN directly via the auxiliary reconstruction loss, the second term trains the full system end-to-end on the final output, and the third is the score matching loss for the \texttt{Score} DNN.
To simultaneously support the one-step inference sampler and the standard generative reverse process, the \texttt{Score} DNN is evaluated twice per step under shared weights: once using $\hat{\VEC{S}}^{\mathrm{cond}}_{1:\ell}\!+\!\sigma_t \VEC{Z}_{1:\ell}$ (the noised \texttt{Cond} DNN output) and once using $\VEC{S}_{1:\ell}\!+\!\sigma_t \VEC{Z}_{1:\ell}$ (the noised clean speech) utilizing the same speech conditions $\{\VEC{C}_i\}$ and the same Gaussian noise.

The limited lookahead introduces the only training-data change relative to~\cite{Lugo2026}.
The \texttt{Cond} DNN receives input frames $\VEC{X}_{1:\ell+N_{\mathrm{L}}}$ and $\VEC{Y}_{1:\ell+N_{\mathrm{L}}}$, i.e., $N_{\mathrm{L}}$ frames beyond the current frame, but all network outputs and the clean speech target $\VEC{S}_{1:\ell}$ remain aligned to frame $\ell$.
During data loading, each training segment is therefore drawn with $N_{\mathrm{L}}$ additional frames appended to the input while the target window stays fixed at $L$ frames, ensuring that future context is available to \texttt{Cond} without leaking into the loss computation.

\section{Experimental Setup}
\subsection{Dataset and Framework}
Our training, validation, and test data setup follows~\cite{Lugo2026}.
In brief, training data $\mathcal{D}_\mathrm{train}$ is synthesized from Interspeech 2025 URGENT Challenge speech and noise sources~\cite{Saijo2025} with quality-based filtering~\cite{Li2025}, combined with the synthetic ICASSP 2023 AEC Challenge training set~\cite{Cutler2024}, totaling roughly \qty{623}{\hour} of training data.
Validation data $\mathcal{D}_\mathrm{val}$ is synthesized from the TIMIT speech corpus~\cite{Garafolo1993}, the ETSI noise database~\cite{ETSI-EG-202-396}, and the Aachen impulse response database~\cite{Jeub2009}.
The test set $\mathcal{D}_\mathrm{test}$ is the blind test set of the ICASSP 2023 AEC Challenge~\cite{Cutler2024}.
Since we do not model delay in our setup and as our \texttt{DiffVQE2(-S)} models operate causally, we apply causal delay compensation via GCC-PHAT~\cite{Knapp1976} to $\mathcal{D}_\mathrm{test}$ prior to evaluation, in contrast to the non-causal \texttt{DiffVQE}~\cite{Lugo2026}.

\subsection{Metrics}
We use the same evaluation suite as in~\cite{Lugo2026}.
AECMOS~\cite{Purin2021} provides echo suppression (Echo) and near-end speech quality (Other) scores across single-talk far-end (STFE), single-talk near-end (STNE), and double-talk (DT) conditions.
DNSMOS~\cite{Reddy2021} (OVRL) gives a non-intrusive overall quality estimate, as well as a speech quality estimate (SIG) and a background noise quality estimate (BAK).
On $\mathcal{D}_\mathrm{val}$, we additionally report PESQ~\cite{ITU-P862}, ESTOI~\cite{jensen2016algorithm}, and the Levenshtein phone similarity ($\text{LPS}=1\!-\!\text{LPD}$, with LPD from~\cite{Pirklbauer2023}) that detects hallucinations specific to generative models.
We also compute the average rank based on standard competition ranking for all non-intrusive metrics (AECMOS, DNSMOS) on $\mathcal{D}_\mathrm{test}$ as a single summary of overall model performance.
Moreover, we conduct crowd-sourced P.808/P.831 subjective listening tests following \cite{ITU-P831,ITU-P808,Cutler2021} using a subset of 50 files per condition (DT/STFE/STNE), providing mean opinion score (MOS) scores.
The subset is selected by iterative greedy removal of files, with a cost that minimizes changes to individual per-metric model ranks and applies a small auxiliary penalty on per-metric mean deviations at each step, such that the average rank order across models is preserved with negligible absolute deviation.
This selection ensures a representative subset whose per-metric distributions closely match the full test set $\mathcal{D}_\mathrm{test}$.

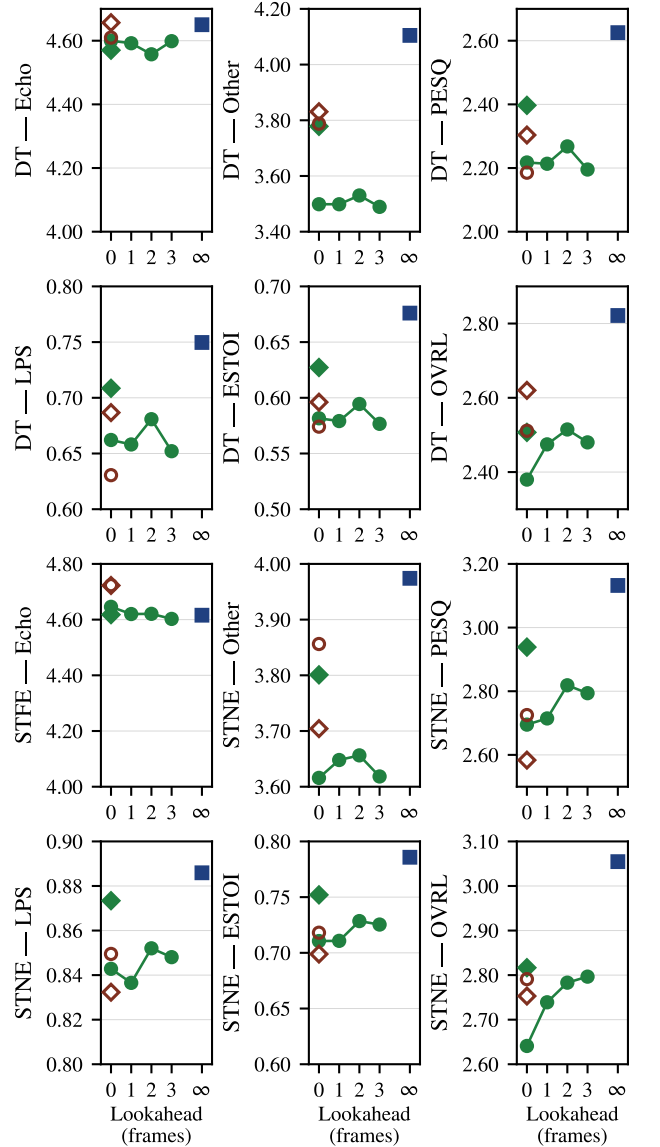
\begin{figure}[!t]
    \centering
    \includegraphics[width=1\columnwidth]{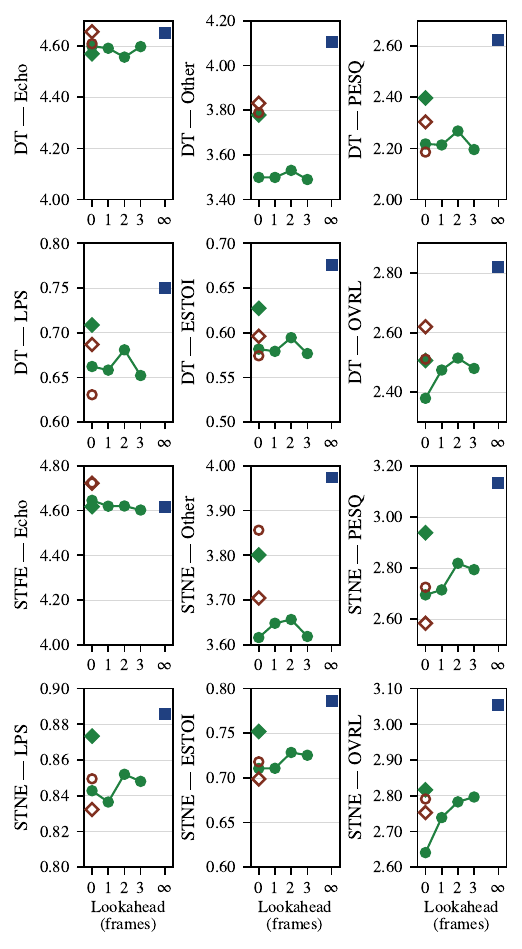}
    \begin{tikzpicture}
\fontsize{9}{9}\selectfont


\coordinate (dCoord) at (3.27, 0.8);
\def\dSize{0.135};
\draw[-, cDarkRed, line width=1.5pt] ($(dCoord)+(-\dSize, 0)$) -- ($(dCoord)+(0, \dSize)$) -- ($(dCoord)+(\dSize, 0)$) -- ($(dCoord)+(0, -\dSize)$) -- ($(dCoord)+(-\dSize, 0)$) -- ($(dCoord)+(0, \dSize)$);
\node[anchor=west, inner sep=0pt] at (3.55cm, 0.8cm)
    {\fontsize{9}{9}\texttt{DeepVQE}~\cite{Indenbom2023}};

\draw[cDarkRed, line width=1.5pt] (5.85cm, 0.8cm) circle (0.097);
\node[anchor=west, inner sep=0pt] at (6.32cm, 0.8cm)
    {\fontsize{9}{9}\texttt{DeepVQE-S}~\cite{Indenbom2023}};

\fill[cBlue] (0.25cm, 0.15cm) rectangle ++(0.25cm, 0.25cm);
\node[anchor=west, inner sep=0pt] at (0.68cm, 0.25cm)
    {\fontsize{9}{9}\texttt{DiffVQE}~\cite{Lugo2026}};

\node[diamond, fill=cGreen, inner sep=2.5pt] at (3.27cm, 0.25cm) {};
\node[anchor=west, inner sep=0pt] at (3.55cm, 0.25cm)
    {\fontsize{9}{9}\selectfont\ttfamily DiffVQE2};

\draw[cGreen, very thick] (5.45cm, 0.25cm) -- ++(0.8cm, 0);
\fill[cGreen] (5.85cm, 0.25cm) circle (3.5pt);
\node[anchor=west, inner sep=0pt] at (6.32cm, 0.25cm)
    {\fontsize{9}{9}\selectfont\ttfamily DiffVQE2-S};

\draw[rounded corners=3pt, gray!40, thick] (0, 0) rectangle (8.6106cm, 1.1cm);
\path[use as bounding box] (0, 0) rectangle (8.6106cm, 1.1cm);
\end{tikzpicture}
    \caption{Influence of lookahead on $\mathcal{D}_\mathrm{val}$ performance in all conditions.}
    \label{fig:val_ablation}
\end{figure}

\subsection{Training Details}
All models are trained on $\mathcal{D}_\mathrm{train}$ resampled to \qty{16}{\kilo\hertz}, using a short-time Fourier transform (STFT) with frame length $512$, frame shift $128$, a square-root Hann window, and frequency bins padded from $K\!=\!257$ to $260$.
To cover single-talk conditions, we remove either the near-end or far-end component in \qty{6.25}{\percent} of samples each, and substitute reverberated near-end speech with the non-reverberated signal in \qty{10}{\percent} of samples to ease the learning target and improve generalization to unseen room impulse response (RIR) characteristics.
Our network topologies build upon~\cite{Lugo2026}.
Our \texttt{DiffVQE2} employs $C_i\!\in\!\{11,16,23,33,50\}$ encoder/decoder channels (same as \texttt{DiffVQE}), with $i$ being the down-/upsampling block index. For details see~\cite{Lugo2026}.
For our smaller \texttt{DiffVQE2-S}, we choose $C_i\!\in\!\{8,16,16,16,32\}$.
GRUs use hidden size $385$ (\texttt{DiffVQE2}) and grouped GRUs ($g\!=\!8$) hidden size $144$ (\texttt{DiffVQE2-S}).
All remaining (diffusion) network hyperparameters, including convolutional kernel sizes and strides, follow~\cite{Lugo2026}.
We investigate lookaheads of $N_\mathrm{L}\in\!\{0,1,2,3\}$ frames.
We train for \qty{500}{\kilo\relax} steps on an \texttt{NVIDIA RTX PRO 6000} with batch size $16$, \qty{8}{\second} random crops, and a three-phase learning rate schedule: linear warmup to $8\!\times\!10^{-4}$ over \qty{7.5}{\kilo\relax} steps, constant through \qty{250}{\kilo\relax} steps, then cosine annealing to $1.6\!\times\!10^{-6}$.
We retrain \texttt{DeepVQE} and \texttt{DeepVQE-S} on $\mathcal{D}_\mathrm{train}$ for the same number of epochs, using the original batch size and learning rate \cite{Indenbom2023}.
The final checkpoint of all models is selected according to the model-individual lowest average rank of metrics on $\mathcal{D}_\mathrm{val}$.

\begin{table*}[!t]
\caption{Model footprint, algorithmic delay, performance on the ICASSP 2023 AEC Challenge test set $\mathcal{D}_\mathrm{test}$. Separate rankings (large/small).}
\label{tab_results}
\footnotesize
\centering
\setlength\tabcolsep{1.4pt}
\renewcommand{\arraystretch}{1.4}
\begin{tabularx}{17.8cm}{l | r r c c | c c c c c c c c c c c|ccccc}
\toprule
\multirow{4}{*}{Method} & \multirow{4}{*}{\rotatebox{90}{\#Param. (M)}} & \multirow{4}{*}{\rotatebox{90}{\#GFLOPS}} & & & & & & & & & & & & & & & & & \\
&&& Alg.  & Look- & DT   & DT    & DT   & DT  & DT  & STFE & STNE  & STNE & STNE & STNE & Avg.
& DT    & DT    & STFE  & STNE & Avg.
\\ 
                        &                           &                          & Delay & ahead & Echo & Other & OVRL & SIG & BAK & Echo & Other & OVRL & SIG  & BAK  & Rank$\downarrow$ 
& MOS E & MOS O & MOS E & MOS O & MOS
\\
& & & & & & & & & & & & & & & & & & & \\
\hline

\texttt{Unprocessed} & - & - & - & - 
& 1.76 & 4.10 & 2.12 & 3.10 & 2.23 & 2.07 & 3.79 & 2.75 & 3.33 & 3.37 & - 
& 2.10              & 2.87              & 2.52              & 3.45              & 2.74\\

\texttt{DiffVQE} \cite{Lugo2026} & 5.1 & 5.37  & $\infty$ & $\infty$ 
& 4.62 & 4.11 & 2.93 & 3.25 & 3.96 & 4.43 & 4.26 & 3.14 & 3.43 & 4.07 & - 
& 3.70              & 3.54              & 3.71              & 3.77              & 3.68\\

\texttt{DeepVQE} \cite{Indenbom2023} & 5.3 & 42.24 & \qty{32}{\ms} & 0 
& 4.64 & 3.84 & 2.75 & 3.11 & 3.86 & 4.37 & 3.93 & 2.97 & 3.31 & 4.03 & 1.2 
& 3.66              & 3.26              & 3.48              & 3.34              & 3.44\\

\texttt{DiffVQE2} & 5.1 & 5.37 & \qty{32}{\ms} & 0 
& 4.48 & 3.76 & 2.67 & 3.00 & 3.86 & 4.46 & 3.88 & 2.98 & 3.28 & 4.02 & 1.7 
& 3.76              & 3.39              & 3.77              & 3.39              & 3.58\\

\hline

\texttt{DeepVQE-S} \cite{Indenbom2023} & 0.7 & 3.12  & \qty{32}{\ms} & 0 
& 4.51             & 3.70             & 2.65             & \textbf{3.02}    & 3.82             & 4.16             & 3.81             & 2.92             & 3.25             & 4.01             & 4.0 
& \textbf{3.62}     & 3.20              & 3.50              & 3.15              & 3.37\\

\texttt{DiffVQE2-S} & 0.7 & 2.57 & \qty{32}{\ms} & 0 
& 4.45             & 3.68             & 2.59             & 2.90             & 3.83             & \underline{4.51} & 3.96             & 2.97             & \underline{3.26} & 4.01             & 4.1 
& 3.44              & 3.10              & 4.03              & \underline{3.64}  & 3.55\\

\multicolumn{1}{c|}{\resizebox{1.2mm}{!}{$\mid$}} & 0.7 & 2.58  & \qty{40}{\ms} & 1 
& 4.49             & \textbf{3.79}    & \underline{2.67} & 2.99             & \underline{3.86} & 4.50             & 4.00             & \textbf{3.00}    & \textbf{3.28}    & \textbf{4.03}    & 2.3 
& 3.45              & 3.16              & 3.93              & 3.52              & 3.52\\

\multicolumn{1}{c|}{\resizebox{1.2mm}{!}{$\mid$}} & 0.7 & 2.59  & \qty{48}{\ms} & 2 
& \underline{4.52} & \textbf{3.79}    & 2.66             & 2.98             & \textbf{3.87}    & \underline{4.51} & \textbf{4.04}    & \underline{2.99} & \textbf{3.28}    & 4.01             & \underline{2.1} 
& 3.45              & \textbf{3.39}     & \underline{4.08}  & 3.57              & \underline{3.62}\\

\multicolumn{1}{c|}{\resizebox{1.2mm}{!}{$\mid$}} & 0.7 & 2.60  & \qty{56}{\ms} & 3 
& \textbf{4.54}    & \underline{3.78} & \textbf{2.68}    & \underline{3.01} & \textbf{3.87}    & \textbf{4.53}    & \underline{4.03} & \textbf{3.00}    & \textbf{3.28}    & \underline{4.02} & \textbf{1.5} 
& \underline{3.61}  & \underline{3.27}  & \textbf{4.10}     & \textbf{3.66}     & \textbf{3.66}\\

\bottomrule
\end{tabularx}
\vspace{-0.5cm}
\end{table*}

\section{Experimental Evaluation and Discussion}

In \autoref{fig:val_ablation}, we report the performance of \texttt{DiffVQE2(-S)} and all baselines on $\mathcal{D}_\mathrm{val}$ across all conditions, as the lookahead $N_\mathrm{L}$ is varied for the small model (\texttt{DiffVQE2-S}).
Reference-free quality is tracked with AECMOS and DNSMOS, and complemented by the intrusive metrics PESQ, LPS, and ESTOI.

\autoref{fig:val_ablation} shows that all investigated models have a very strong echo performance, documented by DT-Echo and STFE-Echo in the range of an excellent 4.6 or even above. Concerning the preservation and fidelity of the near-end speaker's signal, the metrics DT-Other and STNE-Other report advantages of the \texttt{DeepVQE(-S)} baselines over our methods. Analyzing this observation a bit deeper, let's look into instrumental intrusive metrics reflecting speech quality and intelligibility. Only looking onto 0-frames lookahead, our efficient model \texttt{DiffVQE2-S} excels \texttt{DeepVQE-S} in double talk PESQ, LPS, and ESTOI, while in STNE, \texttt{DeepVQE-S} is slightly ahead. \textit{Concerning the larger models, we observe that our \texttt{DiffVQE2} is better than \texttt{DeepVQE} in all intrusive metrics (PESQ, LPS, ESTOI) in both relevant conditions DT and STNE.}

Analyzing the performance of our proposed \texttt{DiffVQE2-S}, we observe that with two frames of lookahead, it turns out to excel \texttt{DeepVQE-S} not only in double talk intrusive metrics, but also in single talk. Concerning DT-OVRL and STNE-OVRL, the two-frames lookahead \texttt{DiffVQE2-S} gets on par with \texttt{DeepVQE-S} on the validation set $\mathcal{D}_\mathrm{val}$.

In \autoref{tab_results}, we report model footprint, algorithmic delay, and reference-free instrumental metrics (AECMOS, DNSMOS) and subjective mean opinion scores (MOS) for our \texttt{DiffVQE2} and \texttt{DiffVQE2-S}, alongside the baselines \texttt{DiffVQE}, \texttt{DeepVQE}, and \texttt{DeepVQE-S}, on the blind test set $\mathcal{D}_\mathrm{test}$.
Moreover, we report the average MOS from the subjective listening test and the average rank for all instrumental metrics, among all small models and among the two causal large models, respectively.

In the upper segment of \autoref{tab_results}, we observe that our novel causal \texttt{DiffVQE2} model loses performance in most instrumental metrics versus its non-causal counterpart \texttt{DiffVQE}. Concerning subjective MOS scores, we see a mixed picture on the isolated MOS scores, resulting in an average MOS loss of 0.1 points (3.58 vs. 3.68). 

Comparing our \texttt{DiffVQE2} to baseline \texttt{DeepVQE} on instrumental metrics, our model reveals some strengths in single-talk conditions, whereas \texttt{DeepVQE} is somewhat better in double talk. These results lead to our \texttt{DiffVQE2} falling short on the average rank for these instrumental metrics (our 1.7 vs.\ 1.2 of the baseline). This, however, is different in subjective metrics: \textit{Here, our \texttt{DiffVQE2} excels the so-far state of the art \texttt{DeepVQE} acoustic echo and noise control model in any of the four MOS metrics, reaching an average of 3.58 vs.\ 3.44 MOS points.} Note that this is achieved under the same algorithmic delay with a slightly smaller model and at 13\% of the computational complexity (5.37 vs.\ 42.24 GFLOPS).

In the lower segment of \autoref{tab_results}, we compare small models of equal size (0.7M) at various GFLOPS and algorithmic delays. Let's start with the most interesting comparison, the baseline \texttt{DeepVQE-S} vs.\ our small \texttt{DiffVQE2-S} model at $0$-frame lookahead: Here, again, the instrumental metrics report certain strengths in double talk for the baseline, whereas our model is a bit better in single-talk conditions. Overall, however, metrics and instrumental ranks (our 4.1 vs.\ 4.0 of the baseline) are very close, with a smaller rank difference than for the large models. While the isolated subjective MOS results provide a mixed picture with the same trends as among the instrumental metrics, \textit{the average MOS of our proposed \texttt{DiffVQE2-S} vs.\ \texttt{DeepVQE-S} is clearly better with 3.55 vs.\ 3.37.} This is achieved at the same model size and algorithmic delay, but at 83\% of the computational complexity (2.57 vs.\ 3.12 GFLOPS).

As many applications allow for a bit more algorithmic delay than just \qty{32}{\ms}, we investigate the potential of various lookaheads. With a few exceptions, the overall trend is that the more lookahead, the higher the performance of our method. We observe that with two frames of lookahead, we achieve a performance comparable to our larger \texttt{DiffVQE2} model, with an average MOS of 3.62 vs.\ 3.58. The price to pay is an algorithmic delay of \qty{48}{\ms} vs.\ \qty{32}{\ms}, but at a model size of only 0.7M (instead of 5.1M) and a computational complexity of only 2.59 GFLOPS (instead of 5.37 GFLOPS).

\section{Conclusions}
In this work, we are the first to propose diffusion-based models for joint acoustic echo and noise control that operate at low algorithmic delay (i.e., causally) and at low computational cost.
Causality is enforced through architectural constraints and an optional model slimming is used to provide models at the two parameter scales of \texttt{DeepVQE} and \texttt{DeepVQE-S} (5.3M / 0.7M parameters) at a smaller respective computational complexity.
We further show that a limited lookahead of our small \texttt{DiffVQE2-S} model improves model performance on par with our larger \texttt{DiffVQE2} model.
In the larger model size regime, our novel efficient low-delay \texttt{DiffVQE2} model excels the so-far state-of-the-art \texttt{DeepVQE} in all four MOS metrics and with an overall MOS of 3.58 vs.\ 3.44 points, at only 13\% of the computational complexity. In the smaller model size regime, our \texttt{DiffVQE2-S} model excels \texttt{DeepVQE-S} with an overall MOS of 3.55 vs.\ 3.37, at 17\% less computational complexity.

\newpage

\bibliographystyle{IEEEbib}
\bibliography{ifn_spaml_bibliography}

\end{document}